\documentclass{elsart}

\usepackage{graphicx}

\usepackage{amssymb}

\begin{document}

\begin{frontmatter}



\title{Probing the nuclear EOS with fragment production}


\author{M. Colonna$^{a,b}$, J.Rizzo$^{a,b}$ ,Ph.Chomaz$^{c}$ 
and M. Di Toro$^{a,b}$}

\address{$^a$LNS-INFN, I-95123, Catania, Italy}
\address{$^b$Physics and Astronomy Dept. University of Catania, Italy}
\address{$^c$GANIL (DSM-CEA/IN2P3-CNRS), F-14076 Caen, France}

\begin{abstract}
We discuss fragmentation mechanisms and isospin transport 
occurring
in central collisions between neutron rich systems at Fermi energies. 
In particular, isospin  effects are analyzed
looking at the correlations between fragment isotopic content 
and kinematical properties. 
Simulations are based on an approximate solution of the Boltzmann-Langevin
(BL) equation. An attempt to solve the complete BL equation, by introducing
full fluctuations in phase space is also discussed. 
\end{abstract}

\end{frontmatter}

\section{Introduction}
\label{}

In the last few years the increased accuracy of the experimental
techniques has renewed interest in nuclear reactions
at Fermi energies. Exclusive measurements, event-by-event analysis,
and a $4\pi$ coverage allow a deeper investigation of the evolution
of the reaction mechanisms with beam energy and centrality. New
insights into the understanding of the nuclear matter equation-of-state ($EOS$)
were gained \cite{DanielewiczSC298}.
In particular, recent experimental and theoretical analyses
were devoted to the study of the properties and
effects of the symmetry term of the $EOS$ (iso-$EOS$)
away from saturation conditions \cite{BaoAnBook01,rep1}.
In particular, 
heavy ion reactions with exotic nuclei at Fermi energies 
can be used to study the
properties of the 
symmetry term at densities below and around the normal value.
In central collisions at 30-50 MeV/A, where the full disassembly of the system
into many fragments is observed, one can study specifically properties of 
liquid-gas phase transitions occurring in asymmetric matter \cite{rep1,mue95,bao197,rep}. For instance, 
in neutron-rich matter, phase co-existence leads to  
a different asymmetry
in the liquid and gaseous phase:  fragments (liquid) appear more symmetric
with respect to the initial matter, while light particles (gas) are 
more neutron-rich.
Hence the analyisis of the isotopic content
of all reaction products, from pre-equilibrium emission to fragments,   
allows to get information on low-density properties of the 
isovector part of the nuclear interaction. 

In recent years, 
the properties of fragments and light clusters emitted in 
systems with different initial asymmetries have been
widely investigated \cite{tsangprl1,tsangprl2,botvina,Ger04,sherry,Kow,EPJA},
looking in particular at   
the production yields
of various isotopes, as obtained in reactions between proton-rich 
and neutron-rich 
systems.

More recently, the study of the isotopic content of pre-equilibrium
emission  has revealed a good sensitivity to the iso-EOS, considering 
the emitted neutron to proton ratio as a function of the kinetic energy
\cite{Famiano}.
Here we extend this type of investigation to fragments.
Correlations between fragment charges and velocities
have been recently observed, providing information on the
interplay between thermal and entrance channel (collective) effects in the 
fragmentation mechanism \cite{EPJA_tab,Indra}. 
Following this line, one can also investigate 
correlations between fragment isotopic content and
kinematical properties, trying to get
a deeper insight into the reaction path and to improve the understanding
of the underlying mechanisms. 
In this way one
can also study more in detail the effects of different EOS's and, in 
particular, of the symmetry energy on 
fragment properties \cite{lionti}. 

The collision dynamics is described on the basis of
approximate treatments of the Boltzmann-Langevin
(BL) approach, i.e. of transport equations including fluctuations.
In particular, we will adopt the Stochastic Mean Field (SMF) method,
where fluctuations are projected onto the ordinary space \cite{Salvo}.

A new method to solve the complete BL equation, by 
implementing the full fluctuations in phase space will be discussed in the last part of the manuscript 
(Section 4).  

\section{The model}
Theoretically the evolution of complex systems  
under the influence of fluctuations can be described
by a transport equation with a fluctuating term, the so-called
Boltzmann-Langevin equation (BLE):
\begin{equation}
{{df}\over{dt}} = {{\partial f}\over{\partial t}} + \{f,H\} = I_{coll}[f] 
+ \delta I[f],
\end{equation}
 where $f({\bf r},{\bf p},t)$ is the one-body distribution function, 
 $H({\bf r},{\bf p},t)$ is the one-body Halmitonian and 
$\delta I[f]$ represents the fluctuating part of the two-body
collision integral \cite{Ayik,Randrup}.

Here we will follow the approximate treatment to the BLE 
presented in Ref.\cite{Salvo}, 
the so-called Stochastic Mean Field (SMF) model, 
that 
consists in the implementation of spatial density fluctuations.

Calculations have been performed using the 
TWINGO code, 
where the test particle
method is used to solve Eq.(1) \cite{TWINGO}.
We adopt a soft EOS, with compressibility modulus $K = 200 MeV$ and,
for the density  ($\rho$) dependence of the symmetry energy, 
we consider two 
representative parameterizations, 
$E_{sym}(\rho,I)/A \equiv C_{sym}(\rho)I^2, 
I \equiv (N-Z)/A$ : 
one showing a rapidly increasing behaviour 
 with density, roughly proportional to $\rho^2$ (asystiff)
and one with a kind of saturation above normal
density (asysoft, $SKM^*$) (see Ref.s\cite{rep1,BaranNPA703} 
for more detail).
The two parameterizations 
obviously cross at normal density. The symmetry energy
at densities below the normal value 
is larger in the asysoft case, 
while above normal density it is higher in the asystiff case.  
Hence in the low-density regime, that is the
region of interest for our analysis, isospin effects are expected to
be stronger in the asysoft case. 





\section{Fragment isotopic properties and correlations}
We will focus on central collisions, $b = 2~fm$, considering symmetric reactions
between systems having three different initial asymmetry: 
$^{112}Sn + ^{112}Sn,^{124}Sn + ^{124}Sn,
^{132}Sn + ^{132}Sn,$ with $(N/Z)_{in}$ = 1.24,1.48,1.64, respectively. 
The considered 
beam energy is 50 MeV/A.
1200 events have been run for each reaction and for each of the two 
symmetry energies adopted. 

The first two reactions, $^{112}Sn + ^{112}Sn$ and $^{124}Sn + ^{124}Sn$
have been widely  investigated both from the experimental 
and theoretical point of view \cite{tsangprl1,BaranNPA703,bao,nnoi}. 
In central collisions, after the initial collisional shock, the system
expands and breaks up into many pieces, due to the development of volume
(spinodal) and surface instabilities.  The formation of  
a bubble-like configuration is observed, where the initial 
fragments are located. 
The average fragment multiplicity is approximately equal to 6 
for the reactions considered here \cite{BaranNPA703}.
Along the reaction path, several nucleons are emitted at the early stage
(pre-equilibrium emission) and/or are evaporated while fragments are formed. 
Primary fragments are identified by applying 
a coalescence procedure to the matter with density larger than 
$\rho_{cut} = 1/5~\rho_0$ (liquid phase).
The remaining particles are considered as belonging to the gas phase.

First, let us briefly
recall  some general features concerning the isotopic content of fragments and
emitted nucleons, as obtained with the two iso-EOS's considered.
In the following we will restrict our analysis to fragments with charge
in the range between 3 and 10 (that we call intermediate mass fragments 
(IMF's) ).
The average N/Z of emitted nucleons (gas phase) and IMF's
is presented in Fig.1 as a function of the initial $(N/Z)_{in}$ 
of the three colliding systems.
\begin{figure}
\vspace{0.9cm}
\begin{center}
\includegraphics[width=6.cm]{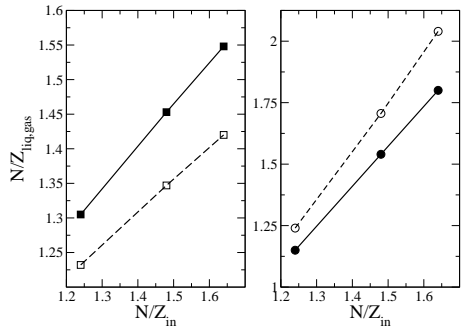}
\caption{The N/Z of the liquid (left) and of the gas (right) phase
is displayed as a function of the system initial N/Z.
Full lines and symbols refer to the asystiff parameterization. Dashed
lines and open symbols are for asysoft. 
} 
\end{center}
\end{figure}
One observes that, generally,   
the gas phase (right panel) is more neutron-rich in the
asysoft case, while IMF's (left panel) are more symmetric. 
This is due to the larger value of the symmetry energy at low density
for the asysoft parameterization 
\cite{BaranNPA703}. 
It is interesting to note that, in the asystiff case, due to the 
rather low value of the symmetry energy,
Coulomb effects dominate and the N/Z
of the gas phase becomes lower than that for IMF's, because protons are
preferentially emitted. 
Now we move to investigate in more detail
correlations between fragment isotopic 
content and kinematical properties. 
The idea in this investigation
is that fragmentation originates from the break-up of a composite source
that expands with a given velocity field.  
Since neutrons and protons experience different forces, 
one may expect a different radial flow for the two species. 
In this case,  the N/Z composition of the source would not be uniform, 
but would depend on the radial distance from the center or mass or, 
equivalently, on the local velocity.
This trend should then be reflected in the fragment asymmetries.  
As a measure of the isotopic composition of the IMF's, we will consider
the sum of neutrons, $N = \sum_i N_i$, and protons, $Z = \sum_i Z_i$, 
of all IMF's in a given kinetic energy range, 
in each event. Then we take the ratio $N/Z$ and we consider
the average over the ensemble of events. 
This observable is plotted in Fig.2 for the three reactions. 
The behaviour observed is rather sensitive to
the iso-EOS. 
For the proton-rich system, the N/Z decreases with the
fragment kinetic energy, expecially in the asystiff case, where the
symmetry energy is relatively low at low density \cite{BaranNPA703}.
In this case, the Coulomb repulsion
pushes the protons towards the surface of the system. Hence, more
symmetric fragments acquire larger velocity.
The same effects are responsible for the proton-rich 
pre-equilibrium emission observed in this case (see Fig.1).  
The decreasing trend is less pronounced in the asysoft case
(right panel) because Coulomb effects on protons 
are counterbalanced by the larger
attraction of the symmetry potential. 
In systems with higher initial asymmetry, the decreasing
trend is inversed, due to the larger neutron repulsion in neutron-rich
systems. Larger slopes are always observed in the asysoft case.  

In conclusion, this analysis reveals the existence of 
significant, EOS-dependent correlations between 
the $N/Z$ and the kinetic energy of
IMF's.  This correlation is linked to the different forces experienced
by neutrons and protons along the fragmentation path, that in turn depend
on the detail of the isovector part of the nuclear interaction. 
This study can be considered as complementary to  
pre-equilibrium emission studies \cite{Famiano,Fam1}. 
A parallel investigation of pre-equilibrium and fragment emissions 
would be very important for 
a cross-check of model predictions 
against experimental observables sensitive to different phases 
of the reaction. 

\section{A new way to implement the full BL equation}
The results discussed above were based on an approximate treatment
of the BL equation, where only fluctuations in the ordinary
space are considered.  This is justified by the fact that
 the multi-fragmentation mechanism is mostly dominated 
by spatial density fluctuations.  However, 
  a more accurate representation of the 
full phase space dynamics would  
improve the description of fluctuations of fragment
kinematical properties, allowing to investigate more sophisticated observables
and  to study the fragmentation path in deeper detail.  

Within the test partice method, that is usually adopted 
to solve numerically the BNV
equation,   
each nucleon is represented by a 
collection of $N_{test}$ test particles, that are propagated according
to the mean-field interaction and hard two-body scattering.
Since collisions are treated stochastically for the
single test particles, the dispersion around the average number of nucleon
collisions is automatically divided by $N_{test}$. 
Hence fluctuations are largely reduced. 
\begin{figure}
\begin{center}
\includegraphics[width=6.cm]{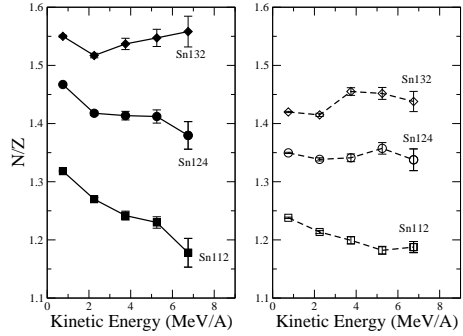}
\caption{
The fragment N/Z (see text) as a function of the kinetic energy. 
Left panel:asystiff;  Rigth panel: asysoft.
} 
\end{center}
\end{figure} 

To overcome this problem,
Bauer et al. \cite{fluct_bauer} 
proposed a method to introduce a correlation between close particles in 
phase-space. 
The method follows the idea, first applied in extended TDHF, 
 of evolving the single-particle density 
including a perturbative mixing of Slater determinants. 
This is realized in the collision 
integral by means of the following procedure: 
The nucleon-nucleon (NN) cross 
section is reduced by a factor $N_{test}$. 
Two test particles $i_1$ and $i_2$ are chosen as 
colliding partners, and will be moved to new positions 
in ${\bf p}$ space 
provided that the corresponding transition probability
is larger than a random number (extracted between 0 and 1),  
as usually done in transport codes using the test particle method. 
If the two particles can collide, i.e. the final positions are not
forbidden by the Pauli blocking, the 
scattering actually involves two ``clouds'' of neighbouring test particles, 
corresponding to two entire nucleons ($2\times N_{test}$ particles),
that are chosen as the particles at closest distance in phase space.

This method is conceptually simple, and moreover the implementation into 
existing transport codes is really straightforward. 
However, since the Pauli-blocking is only checked for the first two
colliding test particles and not for the entire swarm of particles that
are actually moved, at the end  
the fluctuation strength does not reproduce the
expected value for fermionic systems, but approaches instead the classical
value.  

A careful modification of the original 
procedure can considerably improve the results. 
The procedure that we propose can be summarized in the following steps:

\begin{enumerate}

\item The choice of the two colliding partners closely traces the 
standard recipe. If two test particles are allowed to collide, 
moving from the initial positions ${\bf p_1}, {\bf p_2}$ to new
positions, ${\bf p_3}, {\bf p_4}$  corresponding to a rotation by ($\theta,\phi$) with respect to the relative momentum direction, 
two clouds of $N_{test}$ particles are moved. 

\item
 Only particles within a sphere, in coordinate
space,  
around the 
center of mass of the two partners $i_1$ and $i_2$ can belong to the 
clouds. The distance criterion is 
$|\mathbf{r}_i-\mathbf{r}_{CM}(i_1,i_2)|<d_r$, where $d_r$ is a free 
parameter (see later discussion).
Then a grid is introduced in momentum space, the size 
of each cell being $V_{cell}$. 

\item indicating by $I$ and $J$ the cells 
containing the partners $i_1$ and $i_2$, 
we consider the corresponding final cells $I'$ and $J'$ in the 
frame rotated by  ($\theta,\phi$). 
For a given set of initial and final cells, 
the number of test particles that will be actually moved to final states is 
equal to the minimum between the occupation of the initial cells and 
the availability of the final ones $\bar{n'}=(1-n')$:
$$
n_t(I,I';J,J')= \min(n_I,n_J,\bar{n}_{I'},\bar{n}_{J'})
$$

\item Surrounding cells 
are explored with the same prescriptions, until one entire nucleon
is reconstructed around $i_1$ and $i_2$. 
The search procedure is symmetric with respect to the 
center of mass of the two partners.

\item 
A further check on the total momentum of the two clouds is performed, and 
the origin of the grid is eventually slightly displaced in order to have a 
perfect energy and momentum conservation. 

\end{enumerate}
This method involves two parameters, namely the radius of the sphere in 
$\mathbf{r}$ space, $d_r$, and the size of the momentum cells, $V_{cell}$. 
The radius $d_r$ fixes the spatial extension of the nucleon, that in turn 
influences its spreading in momentum space.
Hence this parameter fixes somehow the extension of the nucleon wave packet
in phase-space and generally could affect the transport 
dynamics.  
It can be constrained by physical 
arguments, and in general it depends on physical properties of the 
system, such as its dilution. 
The cell size $V_{cell}$ should be 
small enough to allow an accurate check of the Pauli blocking, but large 
enough to contain a sufficient number of test particles to reduce numerical 
uncertainties. 

\subsection{Check of fluctuations}
We discuss our procedure to build fluctuations in the context of
a system of fermions at equilibrium, at
a given density and temperature, i.e. initialized according to a Fermi-Dirac
distribution. This is a situation that is easily reached in the course
of a nuclear reaction, after the initial collisional shock.   

We emulate a piece of nuclear matter by taking a box with 
periodic boundary conditions.
Since we will focus on fluctuations in momentum space, only one cell is 
present in coordinate space, and all particles can be chosen to collide 
(no restrictions in $\mathbf{r}$ space). 
The size of the box is $l=26\,fm$, and we put $2820$ nucleons, so that 
the density has the saturation value $\varrho_0=0.16\,fm^{-3}$; each nucleon 
is represented by a collection of $500$ test particles. 
 Besides, we do not consider 
any distinction between neutrons and protons, so that one nucleon 
occupies at least a phase-space volume $h^3/4$. 
We initialize the momenta so as to 
reproduce a Fermi-Dirac profile corresponding to a temperature of $5\,MeV$; 
finally, we consider a constant cross section $\sigma=160\,mb$. 
In these calculations the volume $V_{cell}$ corresponds to a cube of size  
$l_s=30\,MeV/c$ (and coincides with $V_p = h^3/(4V_r)$).
Calculations are stopped when the fluctuation variance saturates.  

First of all, we have 
checked that our fluctuating collision integral preserves 
the average evolution of the system, i.e does not change the one-body
density profile. 
Also the average collision rate is in good agreement with 
analytical expectations \cite{Col_old}.

\begin{figure}[htb]
\centering
\includegraphics[scale=0.3]{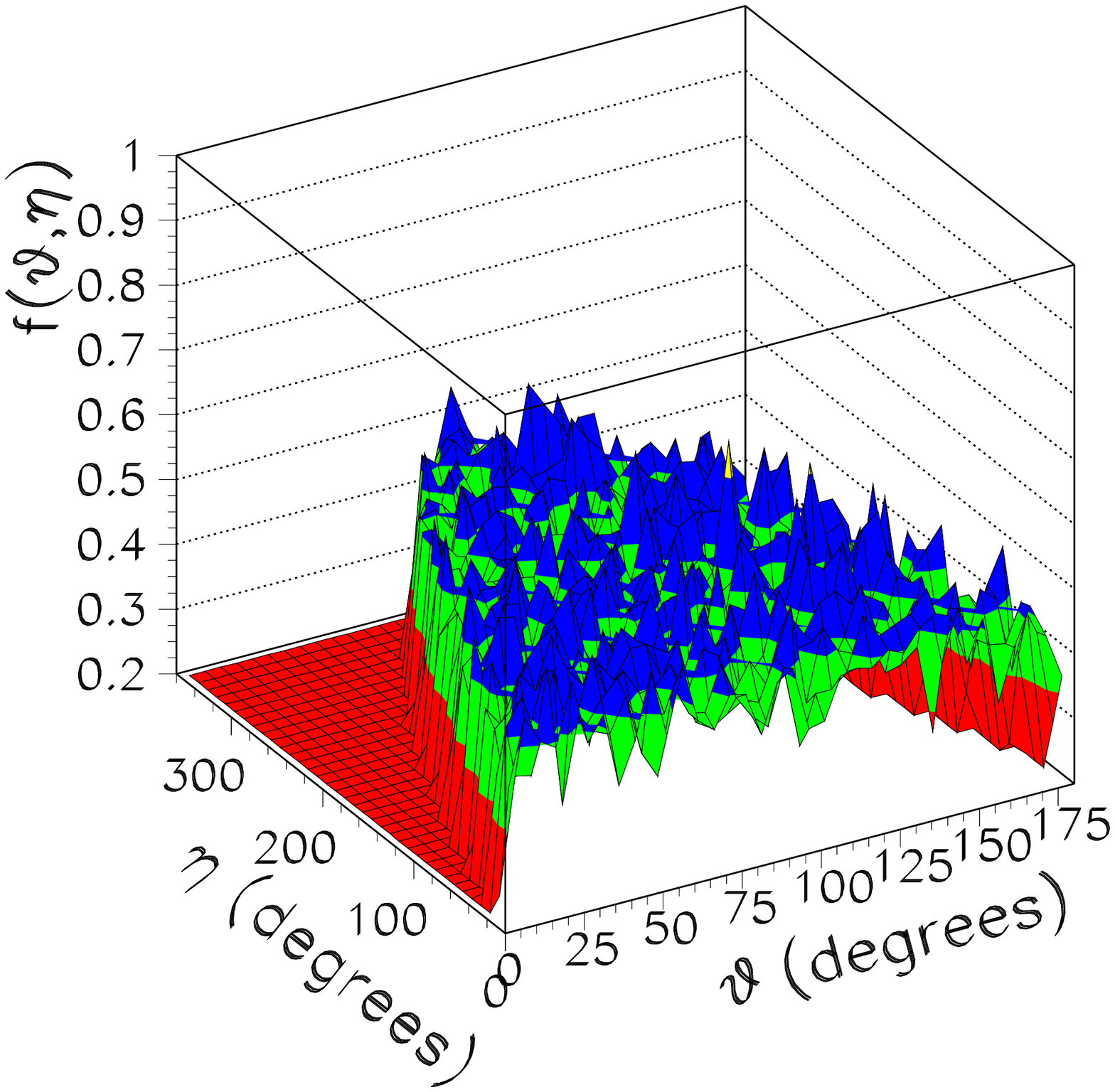} 
\includegraphics[scale=0.3]{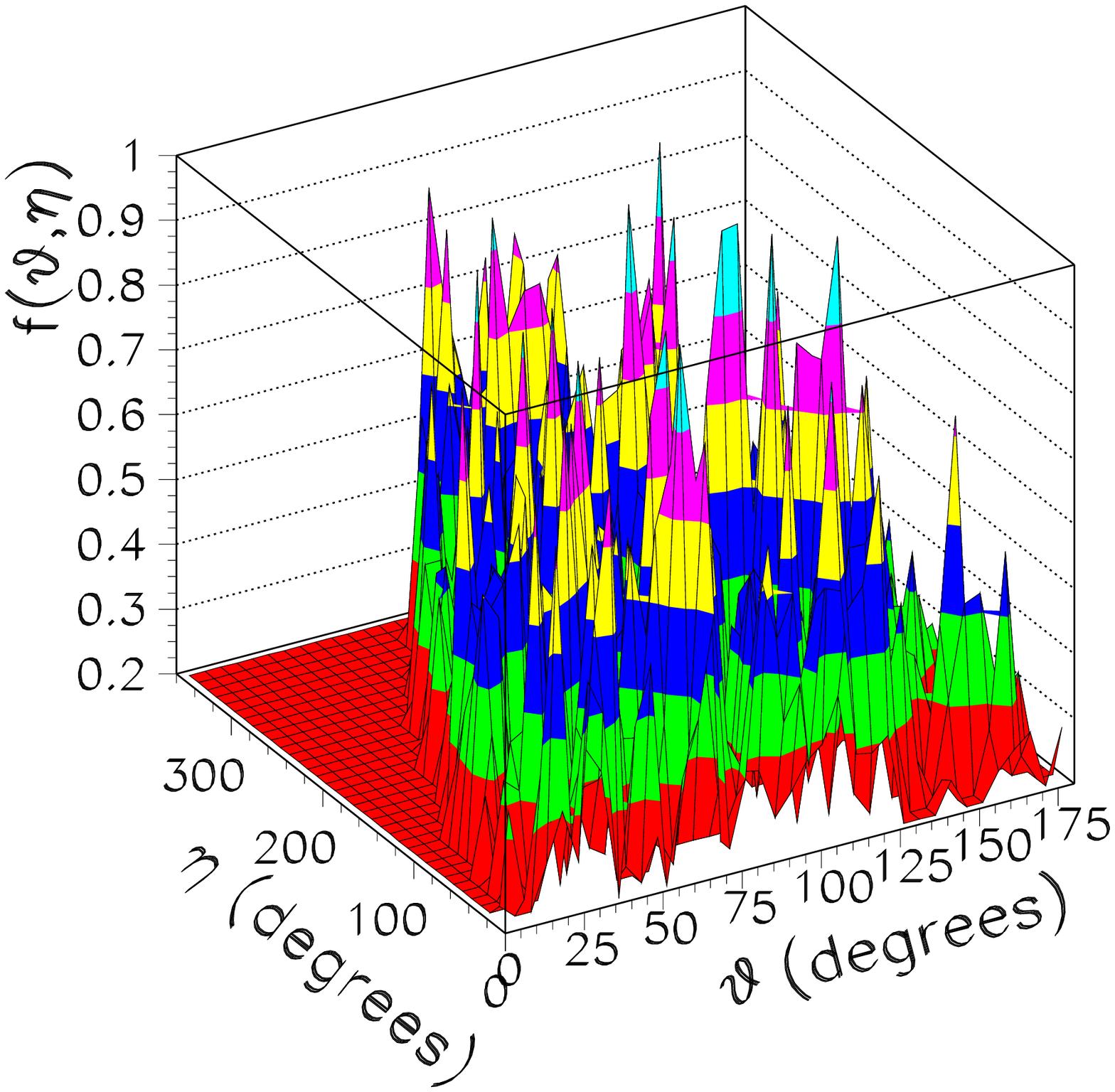}
\caption{Distribution function $f(\theta,\eta)$ at time $t=0$ (left) 
and $t=100\,fm/c$ (right).}
\label{tetaeta}
\end{figure}

One can directly visualize the 
effect of the fluctuating term by selecting a thin region in momentum space 
around the Fermi momentum. This roughly corresponds to look at 
a bi-dimensional system. For this purpose, we adopt 
a new set of coordinates $(dp,\, pd\theta,\, p \sin(\theta)d\phi )$;  
for fixed $p$ and $dp$, the distribution function will depend only on 
two coordinates, namely: 
$$
f(p,\theta,\phi)\rightarrow f(\theta,\eta)
$$
with $d\eta=\sin(\theta)\phi$. 
We choose $p=260\,MeV/c$ (approximately equal to the Fermi momentum) and 
$\Delta p=10\,MeV/c$. In Fig. \ref{tetaeta} we plot the distribution 
function $f(\theta,\eta)$ at two different times (initial and final times). 
At the initial time, $f$ is nearly uniform, and its fluctuations are 
simply due to the numerical noise induced by the finite number of test 
particles. At later times, we observe 
the growth of fluctuations, evidenced by the typical structure with ``peaks 
and holes''. 

In order to check the amplitude of fluctuations built by our 
procedure, we construct the density variance 
in a phase-space cell of volume $V$, $\sigma^2 = <\delta f({\bf p})
\delta f({\bf p})>$. At the Fermi surface, the expected value of the
variance is equal to:  $\sigma^2 = f_{eq}(1-f_{eq})/N_V = 0.25/N_V$, where $N_V$ is the
number of nucleons that can be contained, at most, inside the volume $V$:
$N_V = 4V/h^3$. $f_{eq}$ denotes the average 
equilibrium value of the distribution function that, at the Fermi surface,
is equal to $0.5$.

Since the nucleon wave packet is generally deformed, it extends over volumes
larger than $h^3/4$, hence the correct value of fluctuations
is recovered in the ``continuum limit'', i.e. in large
volumes $V$, as shown in Fig.4. 
The slight deviation from the expected value, $0.25$, is due to some 
difficulties, inherent to the adopted numerical procedure,    
in reconstructing entire nucleons in all collisional events. 
\begin{figure}[htbp]
\vspace{0.8cm}
\begin{center}
\includegraphics[scale=0.3]{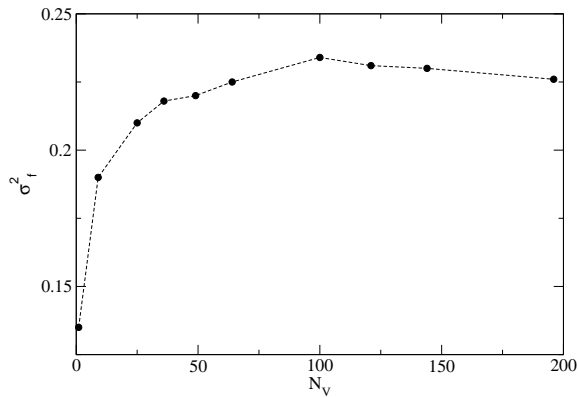}
\end{center}
\caption{Variance of the distribution function (rescaled by $N_V$) as a 
function of the number $N_V$ of nucleons contained in the cell 
considered to construct the fluctuation variance.}
\label{fluct_new}
\end{figure}
  



\section{Conclusions}
We have discussed properties of the fragmentation 
path in central collisions of charge-asymmetric systems, 
that can be 
related to the behaviour of the symmetry energy below normal
density, thus allowing to extract information on fundamental 
quantities of the nuclear interaction. 
We focused on the analysis of correlations between fragment isotopic content
and kinetic energy, performing simulations based on SMF approaches.
An EOS-dependent relation between these two observables is found.
This study also allows one to get a deeper insight 
into the fragmentation mechanism. In fact, the analysis of correlations
between fragment composition and velocity can
be used as a clock of fragment formation and as an indicator of the
underlying dynamics.

A new method to improve the treatment of fluctuations in transport
approaches is also discussed. This is relevant not only for
fragmentation studies, but in general for the dynamical description of 
quantum many-body systems. 


\end{document}